\documentstyle[12pt]{article}
\parskip=\medskipamount
\def\un#1{\underline{#1}}
\def\ba#1{\begin{array}{#1}}
\def\ea{\end{array}}
\def\beq{\begin{equation}}
\def\eeq{\end{equation}}
\def\beqa{\begin{eqnarray}}
\def\eeqa{\end{eqnarray}}

\def\d{\mbox{d}}
\def \CRAS{C.~R.~Acad.~Sc.~Paris}

\begin{document}

\title
{\Large On the exact solutions of the Bianchi IX cosmological model in the
proper time}
\author{\large J.~Springael$^{\dag}$, R.~Conte$^{\ddag}$,
M.~Musette$^{\dag}$\\
\vspace{2mm}
\normalsize \dag\ Dienst Theoretische Natuurkunde\\
\normalsize Vrije Universiteit Brussel\\
\normalsize Pleinlaan 2, B--1050 Brussel, Belgium\\
\vspace{2mm}
\normalsize \ddag\ CEA Saclay, Service de physique de l'\'etat condens\'e\\
\normalsize F--91191 Gif-sur-Yvette Cedex, France}

\date{2 March 1998}

\maketitle

\vspace{60mm}
\begin{abstract}
\noindent 
It has recently been argued that there might exist a four-parameter analytic
solution to the Bianchi IX cosmological model,
which would extend the three--parameter solution of
Belinskii {\it et al.}~to one more arbitrary constant.
We perform the perturbative Painlev\'e test in the proper time variable,
and confirm the possible existence of such an extension.
\end{abstract}
\vfill {\it Regular and chaotic dynamics}, to appear (1998).
\hfill solv-int/9804008
\newpage

\section{Introduction} 

The Bianchi IX cosmological model results from the assumption of
a universe with a homogeneous anisotropic space,
and an important question concerns the 
singularities in the complex $t$-plane of the components $(a^2,b^2,c^2)$ of the
metric tensor
\cite{LandauLifshitzTheorieChamps,DNF,Misner1969a}.
This system is made of three coupled second order nonlinear ordinary
differential equations :
\beq 
2\sigma^2 abc\frac{\d}{\d t}\left(abc\frac{\d (\ln a)}{\d t}\right)
=a^4-(b^2-c^2)^2 
\mbox{\hspace{10mm}and cyclically},
\label{bix1}
\eeq
with the first integral
\beq 
I=\sigma^{2}\left[
a^{2}\frac{\d (b^{2})}{\d t}\frac{\d (c^{2})}{\d t}+
b^{2}\frac{\d (c^{2})}{\d t}\frac{\d (a^{2})}{\d t}+
c^{2}\frac{\d (a^{2})}{\d t}\frac{\d (b^{2})}{\d t}\right]+
a^4+b^4+c^4-2(b^2 c^2+c^2 a^2+a^2 b^2).
\label{eqFirstIntegral}
\eeq
Here $\sigma^2=$ is $+1$ or $-1$ according as the metric is asymptotically
Euclidian or Minkowskian.
The study of these singularities is easier in the
so-called logarithmic time variable $\tau$, 
related to the proper time variable $t$ through the transformation :
\beq
\d t=abc\ \d\tau. 
\label{transfo}
\eeq 
The main result of the Painlev\'e test~\cite{CoGrRa1}--\cite{CoGrRa3}
is a probable chaotic behaviour of the system. 
Moreover, one can prove \cite{MR1998II}
the inexistence of rational first integrals
other than (\ref{eqFirstIntegral}).
As a by-product, this test can also be used to detect all possible
particular analytic solutions,
in addition to the three particular solutions which were already known.

The first exact solution results from the existence of the Euler subsystem
(1750) :
\beqa
\sigma\frac{\d\omega_1}{\d\tau}=\omega_2\omega_3
\mbox{\hspace{10mm}and cyclically}
\label{euler}
\\
\sigma\frac{\d(bc)}{\d t}=a\mbox{\hspace{10mm}and cyclically}
\label{eulerphys}
\eeqa
with $\omega_1=bc$ and cyclically,
reducible to Weierstrass' elliptic equation in the variable 
$\omega_{1}^{2}(\tau)$ :
\vspace{-5mm}
\beqa
\sigma^2\left(\frac{\d(\omega_{1}^{2})}{\d\tau}\right)^2=4\omega_{1}^{2}
(\omega_{1}^{2}+k_2)(\omega_{1}^{2}+k_3).
\label{weier}
\\
\left(\frac{\d(\omega_{1}^{2})}{\d t}\right)^4=16\omega_{1}^{2}
(\omega_{1}^{2}+k_2)(\omega_{1}^{2}+k_3)
\label{weierphys}
\eeqa
The three--parameter solution of system (\ref{euler}), single-valued in $\tau$,
was found by Abel and Jacobi, 
and later in the study of the Bianchi IX system by Belinskii {\it et al.}
\cite{BeGiPaPo1}.

A second subsystem exists, the so-called Darboux system \cite{Da1} :
\beqa
\sigma\frac{\d\omega_1}{\d\tau}=\omega_2\omega_3-\omega_1\omega_2-
\omega_1\omega_3\mbox{\hspace{10mm}and cyclically}
\label{darboux}
\\
\sigma\frac{\d(bc)}{\d t}=a-b-c\mbox{\hspace{10mm}and cyclically}
\label{darbouxphys}
\eeqa
\noindent which has been integrated by Halphen \cite{Ha1,Bu1}. This 
three--parameter solution, single-valued in $\tau$ was rediscovered in the
context of Bianchi IX, by Gibbons {\it et al.} \cite{GiPo1}.
This system is equivalent to the third order Chazy equation of class III
\cite{Ch1}
\beq
\frac{\d^3 y}{\d\tau^3}-2y\ \frac{\d^2 y}{\d\tau^2}+3\left(\frac{\d
y}{\d\tau}\right)^2=0\label{chazy}
\eeq
with $y=-2 (\omega_1+\omega_2+\omega_3)/ \sigma$.

Last, a third particular solution was found by Taub \cite{Ta1}, 
when the metric tensor is axially symmetric (e.g. $b^2=c^2$), 
leading to a four--parameter trigonometric solution.

As to the Painlev\'e analysis, it identifies three possible local behaviours 
of the general solution in the $\tau$ variable 
\begin{enumerate}
\item
a simple pole for $a^2$ and a simple zero for $b^2$ and $c^2$;
\item
a simple pole for $a^2,\ b^2$ and $c^2$;
\item
a regular behaviour for $a^2$ and a double pole for
$b^2$ and $c^2$.
\end{enumerate}

Finally, the only possible new solution isolated by the Painlev\'e
analysis is \cite{LaMuCo1}
an extension of Belinskii's elliptic solution to four parameters,
whose closed form is yet unknown and which we denote for convenience BGPP4.
The finding of this closed form has a strong interest in cosmology,
since it would {\it ipso facto} yield a six-parameter
single-valued solution to the Brans-Dicke Bianchi IX cosmological model
\cite{DeSc1,ScDe1}.

The aim of this paper is to further examine this four--parameter solution.
The connection between the local behaviours in $t$ and those in $\tau$
is established in section \ref{sectionLocal}.
We perform in section \ref{sectionRecovering}
the perturbative Painlev\'e test
\cite{FoPi1,CoFoPi1} on system (\ref{bix1}) in the proper time variable $t$,
in order to see whether or not the existence of the 
four--parameter solution can be denied.

% -------------------------------------------------------------------------
\section{The local study \label{sectionLocal}}

The successive steps are the following. 
First we determine all possible leading behaviours of the Laurent series 
for the dependent variables $a(t), b(t)$ and $c(t)$. 
This is done by substituting respectively 
$a_0 (t-t_0)^{p_1},\ b_0 (t-t_0)^{p_2},\ c_0 (t-t_0)^{p_3}$ into the system
(\ref{bix1}),  and by checking all the different cases for the constants $p_1,\
p_2,\ p_3$,  under the condition that the constants $a_0,\ b_0,\ c_0$ are all
different from zero. 

Then, for each leading behaviour,
we have to look for the so-called indices or resonances. 
These indices can be found by inserting :
\[ 
a(t)=a_0 \chi^{p_1}(1+\varepsilon a_r \chi^r);\ 
b(t)=b_0 \chi^{p_2}(1+\varepsilon b_r \chi^r);\ 
c(t)=c_0 \chi^{p_3}(1+\varepsilon c_r \chi^r);
\]
\noindent with $\chi=t-t_0$, into the system (\ref{bix1}), 
and by taking only those terms which are linear in $\varepsilon$. 
This leads to a coupled system of three linear equations in $a_r,\ b_r,\ c_r$,
whose solution is unique iff the
determinant of the coefficients differs from zero. 
The indices are the zeroes $r$ of this determinant.

If all the leading exponents are positive, 
it is recommended to invert at least one of the dependent variables, 
for example $a(t)\rightarrow a(t)^{-1}$ in order to calculate the indices. 
For a more detailed explanation we refer to \cite{Cargese96}.

If we now apply the scheme explained above, we obtain only three
possibilities, listed in Table \ref{table1}.

The arbitrary values of the two opposite indices 
of the third family (III) immediately imply an infinite amount of 
logarithmic branching, hence probably chaos.

The last row represents the local transformations, after integration of
relation (\ref{transfo}) in which we introduced the leading
behaviours for $a(t),\ b(t)$ and $c(t)$.
\begin{table}[h]
\caption{
Families of singularities of $a$, $b$ and $c$ in the variable $t$.
Notation is $\varepsilon^2=\varepsilon_{i}^{2}=1$.}
\begin{center}
\begin{tabular}{|c|c|c|c|} \hline & (I) & (II) & (III)
\\ \hline
$(p_1,\ p_2,\ p_3)$ & $\left(- \frac{1}{3},\ \frac{1}{3},\
\frac{1}{3}\right)$ & $(1,\ 1,\ 1)$ & $(1,\ 0,\ 0)$
\\ \hline
$(a_0,\ b_0,\ c_0)$ & $(2\varepsilon\sigma b_0 c_0/3,\ b_0,\ c_0)$ &
$(\varepsilon_1,\ \varepsilon_2,\ \varepsilon_3)/(2\sigma)$ & $(a_0,\ b_0,\
\varepsilon b_0)$
\\ \hline
$r$ & $-1,\ 0,\ 0,\ \frac{2}{3},\ \frac{2}{3},\ \frac{4}{3}$ & $-4, -4, -4, -1, 2,
2$ & $-1,\ 0,\ 0,\ 0,\ -2/(\sigma a_0),\ 2/(\sigma a_0)$
\\ \hline
$\tau- \tau_0$ & $(t-t_0)^{2/3}$ & $(t-t_0)^{-2}$ & $\ln (t-t_0)$
\\ \hline
\end{tabular}
\end{center}
\label{table1}
\end{table}

Analogously, Table \ref{table2} gathers the results of
\cite{CoGrRa1}--\cite{LaMuCo1}, who performed the same kind of analysis  
in the $\tau$-variable and for the dependent variables
$A=a^2,\ B=b^2,\ C=c^2$.

The comparison of the two tables immediately shows the
specificity of family (III). 
Indeed, the local transformation between the two times
is different for family (III).

For families (I) and (II),
the local transformation yields the correspondence between the exponents $p_j$
and the indices $r$.
On the contrary, for family (III),
the two local laws 
$\tau - \tau_0=(t - t_0)^{-1}$ and $\tau - \tau_0=\ln (t-t_0)$ 
are incompatible,
and the deep reason for that is the non-Fuchsian nature of family (III) 
in the logarithmic time $\tau$.
\newpage
\begin{table}[h]
\caption{Families of singularities of $A=a^2$, $B=b^2$ and
$C=c^2$ in the variable $\tau$.}
\begin{center}
\begin{tabular}{|c|c|c|c|} \hline & (I) & (II) & (III)\\ \hline
$2(p_1,\ p_2,\ p_3)$ & $(-1, 1, 1)$ & $(-1, -1, -1)$ & $(0, -2, -2)$\\ \hline
$(A_0,\ B_0,\ C_0)$ & $(\sigma,\ B_0,\ C_0)$ & $(\sigma,\ \sigma,\
\sigma)$ & $(A_0,\ \sigma^2/A_0,\ \sigma^2/A_0)$\\ \hline
$r$ & $-1, 0, 0, 1, 1, 2$ & $-1, -1 -1, 2, 2, 2$ & $-1, 0, 0, 2$\\
\hline
$\tau- \tau_0$ & $(t-t_0)^{2/3}$ & $(t-t_0)^{-2}$ & $(t-t_0)^{-1}$ \\ \hline
\end{tabular}
\end{center}
\label{table2}
\end{table}

% -------------------------------------------------------------------------
\section{Recovering the exact solutions \label{sectionRecovering}}

Let us now perform the perturbative Painlev\'e analysis
\cite{FoPi1,CoFoPi1} in order, mainly, to test the existence of the assumed
four-parameter extension to the BGPP solution. 

Each of the three families is separately a local representation of the
general solution,
so each family must describe every closed form particular solution.

For the family (I),
we have checked the absence of any movable logarithm,
so none of the known closed form particular solutions in $\tau$ is excluded.

For the family (III),
the matrix $P(r)$ of the usual recurrence relation
\beq
\forall r\ P(r) \pmatrix{a_r \cr b_r \cr c_r \cr}
+ Q_r = 0
\eeq
is
\beq
P(r)=2 b_0^3 \pmatrix{
a_0 b_0 \sigma^2 r (r+1) & a_0^2 \sigma^2 r & a_0^2 \varepsilon \sigma^2 r \cr
0                        & a_0^2 \sigma^2 r^2 -2 & 2 \varepsilon           \cr
0                        & 2         & \varepsilon (a_0^2 \sigma^2 r^2 -2) \cr}.
\eeq
So the number of movable logarithms is :
2 (from the two irrational indices),
plus 1
 (from the difference between the multiplicity 3 and dim(Ker($P(r)))=2$ at index
$r=0$). When one returns to the $\tau$ variable,
the local transformation $\tau - \tau_0=\ln (t-t_0)$ suppresses one of these three
logarithms,
leaving open the possibility of a four-parameter solution,
single valued in the $\tau$ variable.

Let us now process the family (II),
by expanding $a,b,c$ in the double series

\beq 
a=\sum_{n=0}^{+ \infty}\varepsilon^n\sum_{j=n\rho}^{+ \infty}a_{j}^{(n)}t^{j+p}
 =\sum_{n=0}^{+ \infty}\varepsilon^n a^{(n)},
\mbox{\ \ and cyclically},
\label{dubser}
\eeq

\noindent where $\rho$ stands for the lowest index $\rho=-4$.

Due to the invariance under permutation,
it is convenient \cite{LaMuCo1} to introduce the seven new variables 
$(X, Y, Z, U, V, W, T)$

\beq
\left\{
\ba{l} a_{2}^{(0)}=-\sqrt{\frac{2}{3}}Y+\frac{1}{\sqrt{3}}Z\\
b_{2}^{(0)}=\frac{1}{\sqrt{2}}X+\frac{1}{\sqrt{6}}Y+\frac{1}{\sqrt{3}}Z\\
c_{2}^{(0)}=-\frac{1}{\sqrt{2}}X+\frac{1}{\sqrt{6}}Y+\frac{1}{\sqrt{3}}Z
\ea
\right.
\eeq

\beq
\left\{
\ba{l} a_{-4}^{(1)}=-\sqrt{\frac{2}{3}}V+\frac{1}{\sqrt{3}}W\\
b_{-4}^{(1)}=\frac{1}{\sqrt{2}}U+\frac{1}{\sqrt{6}}V+\frac{1}{\sqrt{3}}W\\
c_{-4}^{(1)}=-\frac{1}{\sqrt{2}}U+\frac{1}{\sqrt{6}}V+\frac{1}{\sqrt{3}}W
\ea
\right.
\eeq

\beq a_{-1}^{(1)}+b_{-1}^{(1)}+c_{-1}^{(1)}=T
\eeq

\noindent where the six arbitrary variables are :
$(a_{-4}^{(1)},b_{-4}^{(1)},c_{-4}^{(1)})$, two out of
$(a_{2}^{(0)},b_{2}^{(0)},c_{2}^{(0)})$, and one out of
$(a_{-1}^{(1)},b_{-1}^{(1)},c_{-1}^{(1)})$.
The necessary conditions for the absence of movable branching are the 
following.

\un{Order zero : $n=0$}

\noindent We obtain at $(n, j)=(0, 2)$ the single condition $Z=0$,
i.e.~$a_{2}^{(0)}+b_{2}^{(0)}+c_{2}^{(0)}=0$. 
Parity implies 
$a_{2j+1}^{(0)}=b_{2j+1}^{(0)}=c_{2j+1}^{(0)}=0$ with $j=0,\ 1,\
\dots$.
\vspace{3mm}

\un{Higher orders : $n \ge 1$}

At first order one must distinguish three cases,
and examine them separately at higher orders.

%                                                                     CASE 1
{\bf Case 1} : $X=Y=0$

Because $a_{-4}^{(1)},\ b_{-4}^{(1)},\ c_{-4}^{(1)}$ and for instance 
$a_{-1}^{(1)}$ are arbitrary, one can set for $n>1$ $a_{-4}^{(n)},\
b_{-4}^{(n)},\ c_{-4}^{(n)}\mbox{ and }a_{-1}^{(n)}$ equal to zero; 
the same can been done for $b_{2}^{(n)}$ and $c_{2}^{(n)}$ with $n>0$.
Each series $a^{(n)}$ then terminates

\[
\forall n :\ \ a^{(n)}=t\sum_{j=0}^{n}a_{3j-4n}^{(n)}t^{3j-4n}\ \ \mbox{ with }\ 
a_{-1}^{(1)}=b_{-1}^{(1)}=c_{-1}^{(1)}=\frac{T}{3}.
\]
and we have checked the absence of any logarithm up to 
perturbation order $n=8$. 

This results in a four--parameter Laurent series depending on
$U, V, W, T$. 
The three--parameter solution of Belinskii {\it et al.}~corresponds to $W=0$,
as can be checked by 
substituting the series (\ref{dubser}) into the Euler system (\ref{eulerphys}).

%                                                                     CASE 2
{\bf Case 2} : $(3V^2-U^2)U=0,\ (3Y^2-X^2)X=0$.

No additional no--log conditions are encountered at higher orders $n$, 
and this case isolates 
the three Taub reductions : $b=c\neq a$ and cyclically.

%                                                                     CASE 3
{\bf Case 3} : $U=V=0$.

At second order we find the only additional condition $W=0$, which implies :
\\
$a_{j}^{(n)}=b_{j}^{(n)}=c_{j}^{(n)}=0$ with $j<0, n=1,\ 2,\
\dots $ except for
$a_{-1}^{(1)}=b_{-1}^{(1)}=c_{-1}^{(1)}=T/3$.

This represents a three--parameter solution depending on $X, Y, T$,
which is precisely the Halphen solution to the Darboux system
(\ref{darbouxphys}), as can be checked by direct substitution.

The above results are summarized in Table \ref{table3},
together with their interpretation.

\begin{table}[h]
\caption{Results of the perturbative Painlev\'e test.}

\begin{center}
\begin{tabular}{|c|c|c|c|} \hline order $(n)$ & \multicolumn{3}{|c|}{ }\\ \hline
$n=0$ & \multicolumn{3}{|c|}{$Z=0$}\\ \hline \vspace{-2mm}
 & & $\ (3V^2-U^2)U=0$ & \\ \vspace{-3mm}
$n=1$ &$X=Y=0$ & & $U=V=0$\\
 & & \hbox{\hspace{-80mm}\vbox{$\ (3Y^2-X^2)X=0$\vspace{0.5mm}}\hspace{-80mm}} & \\
\hline  
$n=2$ &  no conditions & $(b-c) (c-a) (a-b) = 0$ &  $W=0$\\ \hline\hline
arbitrary parameters& $U, V, W, T$ & $X, U, W, T$ & $X, Y, T$\\ \hline
identification & BGPP4? & Taub &  Halphen \\ \hline
\end{tabular}
\end{center}
\label{table3}
\end{table}
\vspace{7mm}

\section{Conclusion}

We have applied the pertubative Painlev\'e analysis to
the Bianchi IX cosmological model in the proper time
variable and connected the results to the already
existing ones in the logarithmic time variable.  The
main result is the absence, checked to a high
perturbation order, of any obstacle to the existence of
a four--parameter extension to the solution of Belinskii
{\it et al.}, which would be single valued in the
logarithmic time.

However the challenge still remains to obtain the
closed--form expression of this four--parameter
extension.

\section*{Acknowledgement}

This joint work was made possible by the Tournesol grant
T 95/004. M.~M.~and J.~S.~acknowledge the financial
support extended by  Flanders's Federale Diensten voor
Wetenschappelijke, Technische en Culturele
Aangelegenheden in the framework of the IUAP III no.~9.


\begin{thebibliography}{99}

\bibitem{LandauLifshitzTheorieChamps} Landau~L~D and Lifshitz~E~M 1971 
{\it The classical theory of fields}
(Pergamon Press, Oxford, Third edition and higher).

\bibitem{DNF} Dubrovin~B~A, Novikov~S~P and Fomenko~A~T 1979 
{\it G\'eom\'etrie contemporaine, M\'ethodes et applications} (Nauka, Moscow).
% bbl math Orsay

\bibitem{Misner1969a} Misner~C 1969 
{\it Phys.~Rev.~Lett.}~{\bf 22} 1071--1074.

\bibitem{CoGrRa1} Contopoulos~G, Grammaticos~B and Ramani~A 1993
{\it J.~Phys.~A: Math.~Gen.} {\bf 25} 5795--5799.
 
\bibitem{CoGrRa2} Contopoulos~G, Grammaticos~B and Ramani~A 1994 
{\it J.~Phys.~A: Math.~Gen.} {\bf 27} 5357--5361.
 
\bibitem{LaMuCo1} Latifi~A, Musette~M and Conte~R 1994 
{\it Phys.~Lett.~A} {\bf 194} 83--92.

\bibitem{CoGrRa3} Contopoulos~G, Grammaticos~B and Ramani~A 1995 
{\it J.~Phys.~A: Math.~Gen.} {\bf 28} 5313--5322.
 
\bibitem{MR1998II} J.~J.~Morales-Ruiz and J.-P.~Ramis,
Galoisian obstructions to integrability of Hamiltonian systems II,
20 pages, preprint MA2-IR-98-0004 (19 January 1998).

\bibitem{BeGiPaPo1} Belinskii~V~A, Gibbons~G~W, Page~D~N and Pope~C~N 1978 
{\it Phys.~Lett.~A} {\bf 76} 433--435.
% Three-parameter elliptic solution

\bibitem{Da1} Darboux~G 1878 
{\it Annales scientifiques de l'\'Ecole normale sup\'erieure} {\bf 7} 101--150.
% $(u2 + u3)' = u2 u3$ \'eq.~(124) page 149. Int\'egr\'e en 1881 par Halphen
 
\bibitem{Ha1} Halphen~G~H 1881 
{\it\CRAS}\ {\bf 92} (1881) 1101--1103
% 2 u1' = u1 (u2 + u3) - u2 u3, solution generale (invariants elliptiques)
 
\bibitem{Bu1} Bureau~F~J 1987 
{\it Bulletin de la Classe des Sciences} {\bf LXXIII} 335--353.

\bibitem{GiPo1} Gibbons~G~W and Pope~C~N 1979 
{\it Commun.~Math.~Phys.} {\bf 66} 267--290.

\bibitem{Ch1} Chazy~J 1911 
{\it Acta Math.} {\bf 34} 317--385.

\bibitem{Ta1} Taub~A~H 1951 
{\it Annals of Math.} {\bf 53} 472--490.
% Bianchi IX. Four-parameter trigonometric solution

\bibitem{DeSc1} Demaret~J and Scheen~C 1996 
{\it J.~Phys.~A: Math.~Gen.} {\bf 29} 59--76.
% 7-dim, log time

\bibitem{ScDe1} Scheen~C and Demaret~J 1996 
{\it Class.~Quantum Grav.} {\bf 13} 1909--1930.
% 8-dim, log time, extension of BKL similar to that in B IX

\bibitem{FoPi1} Fordy~A~P and Pickering~A 1991 
{\it Phys.~Lett.~A} {\bf 160} 347--354.
 
\bibitem{CoFoPi1} Conte~R, Fordy~A~P and Pickering~A 1993 
{\it Physica D} {\bf 69} 33--58.

\bibitem{Cargese96} R.~Conte (ed.),
{\it The Painlev\'e property, one century later}, 
about 800 pages,
CRM series in mathematical physics (Springer, Berlin, 1998).
% Carg\`ese, 3--22 June 1996
R.~Conte,
The Painlev\'e approach to nonlinear ordinary differential equations,
1--112, solv-int/9710020.
%M.~Musette,
%Painlev\'e analysis for nonlinear partial differential equations,
%1--54, solv-int/9804003.
 
\end{thebibliography}
\end{document}